\documentclass[useAMS,usenatbib]{mn2e}

\usepackage{graphicx}
\newcommand{\mm}{\,$\umu$m}

\title[Comparative analysis of Seyfert galaxies]{A Mid-IR comparative analysis of the Seyfert galaxies NGC~7213 and NGC~1386}
\author[Daniel Ruschel-Dutra et al.]{Daniel Ruschel-Dutra$^{1}$\thanks{E-mail:
daniel.ruschel@ufrgs.br (AVR)}, Miriani Pastoriza$^{1}$, Rog\'{e}rio Riffel$^{1}$, \and Dinalva A. Sales$^2$ and Cl\'{a}udia Winge$^3$\\
$^{1}$Departamento de Astronomia, Instituto de F\'{i}sica da Universidade Federal do Rio Grande do Sul\\$^2$Department of Physics, Rochester Institute of Technology\\$^3$Gemini Observatory}
\begin{document}

\date{\today}

\pagerange{\pageref{firstpage}--\pageref{lastpage}} \pubyear{2012}

\maketitle

\label{firstpage}

\begin{abstract}

New Gemini mid-infrared spectroscopic observations together with Spitzer Space telescope archival data, are used to study the properties of the dusty torus and circumnuclear star formation in the active galaxies NGC~7213 and NGC~1386. Our main conclusions can be summarised as follows. Polycyclic aromatic hydrocarbon (PAH) emission is absent in the T-ReCS nuclear spectra but is ubiquitous in the data from Spitzer at distances above 100 pc. Star formation rates surface densities are estimated from the 12.8{\mm} [Ne{\sc ii}] line strengths leading to values close to 0.1M$_\odot\,\,{\rm yr}^{-1}\,\,{\rm kpc}^{-2}$. Analogous estimates based on photometric fluxes of IRAC's 8{\mm} images are higher by a factor of almost 15, which could be linked to excitation of PAH molecules by older stellar populations. T-ReCS high spatial resolution data reveal silicate absorption at $\lambda$ 9.7{\mm} in the central tens of parsecs of the Seyfert~2 NGC~1386, and silicate emission in the Seyfert~1 galaxy NGC~7213. In the case of NGC~1386 this feature is confined to the inner 20 pc, implying that the silicate might be linked to the putative dusty torus. Finally, by fitting CLUMPY models to the T-ReCS nuclear spectra we estimate the torus physical properties for both galaxies, finding line of sight inclinations consistent with the AGN unified model.

 \end{abstract}

\begin{keywords}
active galactic nuclei -- interstellar medium -- star formation.
\end{keywords}

\section{Introduction}

The unified model for active galactic nuclei \citep[AGN; ][]{Antonucci1985,Antonucci1993} states that Seyfert~1 and Seyfert~2 (Sy1 and Sy2) galaxies are intrinsically the same object, only viewed from different angles. It is proposed that the line of sight to the type 2 objects happens to cross a highly absorptive medium, supposed to be distributed in a toroidal structure. Some attempts to describe the torus emission considered it as a continuous density distribution \citep[e.g.][]{Pier1992,Granato1997,Siebenmorgen2004,Fritz2006}, however, \citet{Krolik1988} proposed that for dust grains to survive in the torus environment they should be arranged in clumpy structures. Solid dust particles, like silicate and graphite grains, are thought to play a major role in obscuring the AGN. For instance, \citet{Hao2007} showed that quasars and Sy1 are predominantly characterised by silicate in emission at 9.7{\mm} while Sy2 galaxies present weak silicate absorption. While the sublimation of graphite grains creates IR emission at $\lambda\ge1\mu$m, changing the slope of the continuum, the $\sim$9.7$\mu$m feature observed in emission/absorption is attributed to silicate grains \citep[e.g.][]{Barvainis1987,Pier1992,Granato1994,Siebenmorgen2005,Fritz2006,Rodriguez-Ardila2006,Riffel2006,Riffel2009,Elitzur2012,Feltre2012}. Models that assume a clumpy torus simultaneously provide a natural attenuation to UV/optical emission and are able to predict both instances of the silicate feature \citep[e.g.][]{Nenkova2002,Nenkova2008,Nenkova2008a,Honig2006}.

Analysis of the mid-infrared (MID-IR) spectra is of paramount importance to probe the physical processes in regions of elevated optical extinction such as the torus. Besides the already mentioned silicate feature, this spectral window is rich in emission features of molecular hydrogen, polycyclic aromatic hydrocarbons (PAH) and fine structure ionic transitions of which [Ne{\sc ii}] 12.8{\mm}, [Ne{\sc iii}] 15.5{\mm} and [S{\sc iv}] 10.5{\mm} are the most notable examples \citep{Sturm2000,Weedman2005,Wu2009,Gallimore2010,Sales2010}. Of particular importance are the PAH bands at 7.7{\mm} and 11.2{\mm} as they can be used to estimate the nature of the ionising source \citep{Smith2007a,Sales2010}.



To advance our understanding of the interplay between nuclear activity, torus properties and the interstellar medium of the host galaxy, we address here the mid infrared (MIR) properties of two galaxies of contrasting characteristics, namely the Sy2 NGC~1386 and the Sy1 NGC~7213.  
In this paper we present new spectral data from the central regions of NGC~1386 and NGC~7213, acquired with the Thermal Region Camera and Spectrograph (T-ReCS) attached to the Gemini South Telescope.


NGC~1386 is a nearly edge on galaxy with a morphological classification of Sa or S0 depending on the authors \citep{Tully1988,Weaver1991,DeVaucouleurs1964,Sandage1978} and a nuclear activity type Sy2 \citep{Weaver1991}. X-Ray spectral data suggest that this galaxy is Compton-thick, having a column density of $N_{\rm H} \ge 10^{24}\,\,{\rm cm}^{-2}$ \citep{Maiolino1998,Levenson2006}. Optical studies of NGC~1386 have shown the presence of spiral arms and a decoupling between the kinematical and optical nuclei \citep{Weaver1991,StorchiBergmann1996,Rossa2000} which can be interpreted as additional evidence for heavy obscuration of the AGN. The H$\alpha$ image of this galaxy shows extended emission in the form of spiral arms (see fig 2 in \citet{StorchiBergmann1996}). Star formation is also reported for this galaxy \citep{Weaver1991,StorchiBergmann1996}. In addition the soft X-ray spectra of this galaxy is well fitted by two thermal model similar to what have been observed in star burst galaxies \citep{LaMassa2012b}.


NGC~7213 is a Sy1 of morphological type Sa viewed at a face-on angle. The broad components of the H$\alpha$ and H$\beta$ motivate the Sy1 classification, on the other hand low ionisation lines like 6300\AA [O{\sc i}] are relatively strong \citep{Filippenko1984}. In the same paper the authors attribute this low ionisation characteristic to high electron densities ($n_e \sim 10^7$) in the photoionized gas. \citet{StorchiBergmann1996} have identified H$\alpha$ emission from the nucleus and a ring of H{\sc ii} regions at a radius of 20\arcsec from the nucleus, (see fig. 6 in \citet{StorchiBergmann1996}), and also noted a change in the profile of the H$\alpha$ in relation to the observations of \citet{Filippenko1984}. X-ray observations do not show Compton reflection, leading to the conclusion that the Fe K$\alpha$ line might be produced in a Compton-thin torus \citep{Bianchi2003}.

Throughout this paper we adopt a Hubble constant $H_0 = 73$ Km/s/Mpc and the redshifts from \citet{StorchiBergmann1996} as the distance indicator. The redshift $z=2.99\times10^{-2}$ for NGC~1386 leads to a spatial scale of 58 pc/\arcsec and a distance of 11.9 Mpc. For NGC~7213 we have $z=5.84\times10^{-2}$ resulting in a distance of 23.6 Mpc and a spatial scale of 114 pc/\arcsec.
   
We also investigate larger spatial scale structures in these galaxies using archival data from Infrared Array Camera (IRAC) and Infrared Spectrograph (IRS) both aboard the Spitzer Space Telescope. In \S 2 we present a detailed description of the data and reduction processes. \S 3 contains a discussion of the star formation in the host galaxies. Results from the newly acquired T-ReCS data and the properties of the putative dusty tori are discussed in \S 4. Our main conclusions are presented in \S 5.

\section{Observations and Data Reduction}

To build a full picture of the physical process that dominate the MID-IR emission of these two galaxies we used image and spectra from three different instruments: (i) the Infrared Array Camera \citep[IRAC,][]{Fazio2004}; (ii) the Infrared Spectrograph \citep[IRS,][]{Houck2004}, both aboard the Spitzer Space Telescope; (iii) the Thermal-Region Camera and Spectrograph \citep[T-ReCS,][]{Telesco1998,DeBuizer2005} at the Gemini-South Observatory. The latter data set is part of a systematic programme to study the dust and molecular gas in the nuclear region  of Compton thick galaxies (GS-2012A-Q-7, PI Dinalva A. Sales). Spitzer data were retrieved from the public archives. Each set of data is individually discussed in the following subsections.

\subsection{Spitzer - IRAC}
\label{sec.irac}

Spitzer's IRAC is a MID-IR camera capable of simultaneously imaging an area of 5.2\arcmin $\times$ 5.2\arcmin \, in four different filters, with central wavelengths of 3.6{\mm}, 4.5{\mm}, 5.8{\mm} and 8{\mm}, and wavelength intervals 3.2{\mm} to 4.0{\mm}; 4.0{\mm} to 5.0{\mm}; 5.0{\mm} to 6.4{\mm} and 6.4{\mm} to 9.3{\mm} respectively. The data for NGC~1386 were acquired in December 15, 2004, still during the Spitzer's cryogenic mission (program ID: 3269, PI: Jack Gallimore), and at May 9, 2005, for NGC~7213. We only used the post basically calibrated data (post-bcd) from filters 3.6{\mm} and 8{\mm} produced by pipeline number S18.18.0.
 
\begin{figure}
\begin{center}
\includegraphics[width=\columnwidth]{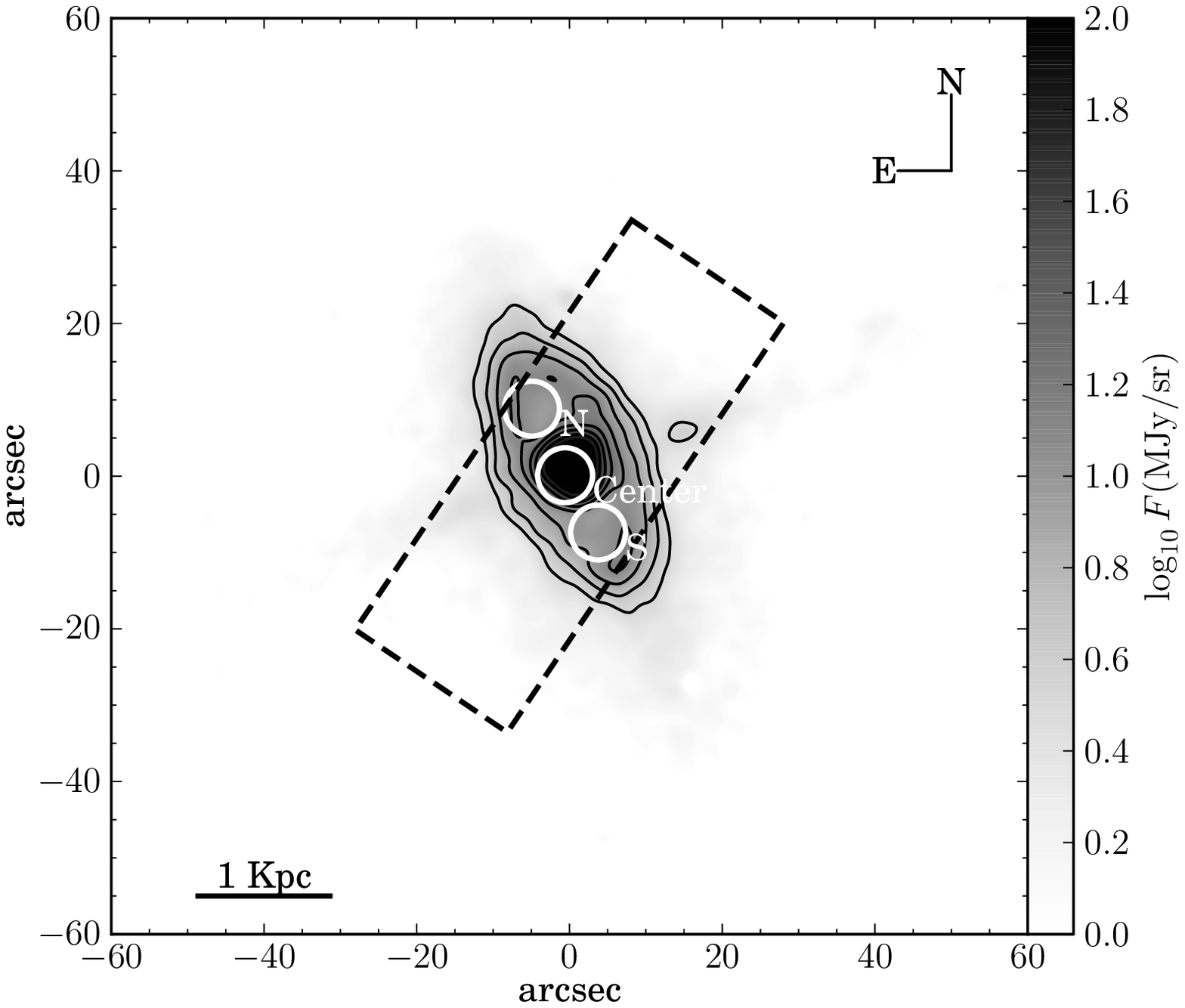}
\includegraphics[width=\columnwidth]{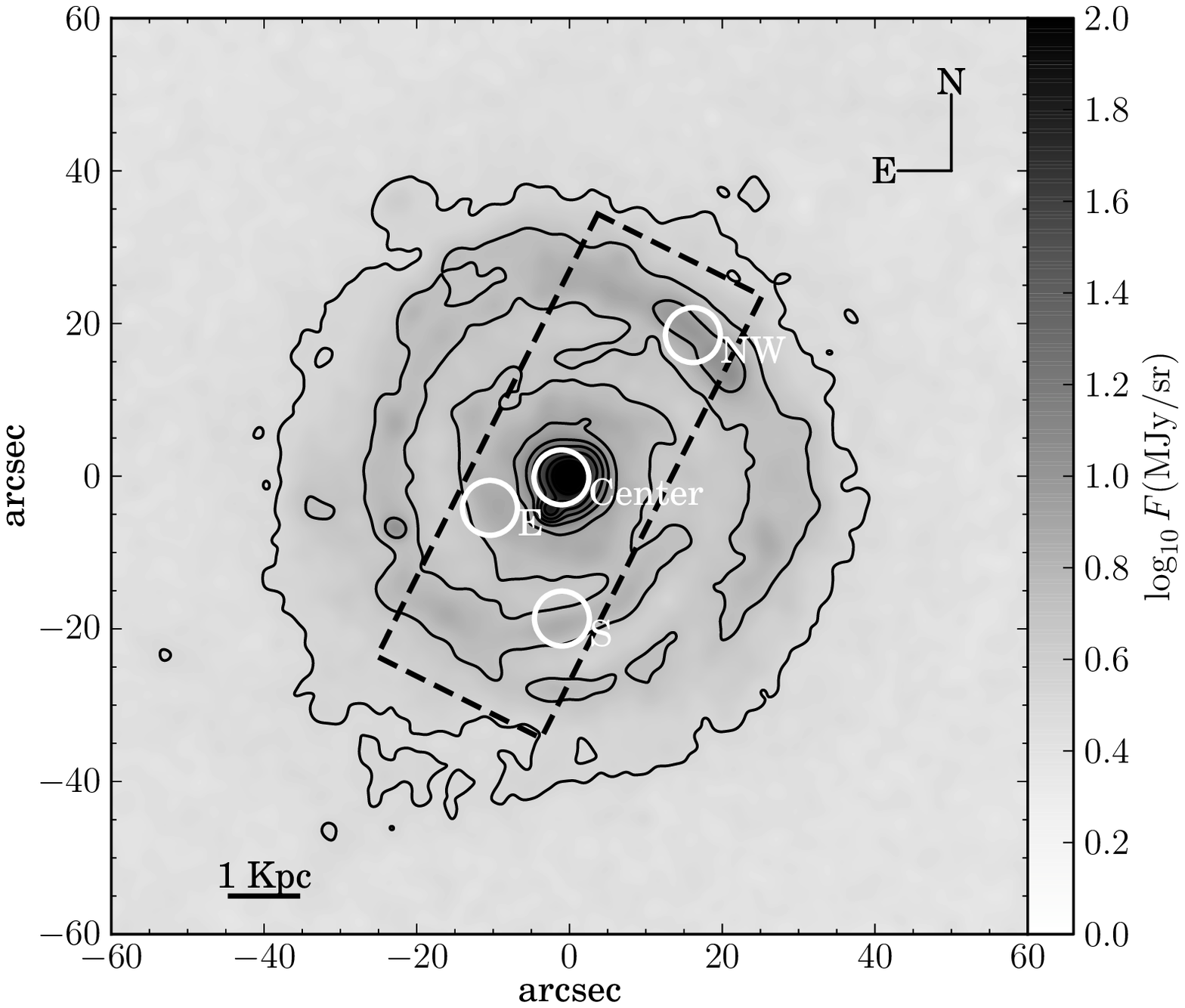}
\end{center}
\caption{IRAC images resulting from the subtraction of the scaled 3.6{\mm} filter from the 8{\mm} filter (refer to section \ref{sec.irac}). The images from top to bottom correspond to NGC~1386 and NGC~7213 respectively, and are centred in the optical nucleus of the galaxies. Circles shown in white correspond to the unresolved nucleus and enhanced PAH emission, as described in the main text. The contours are intervals of 0.2 in $\log_{10} F_\nu (MJy/sr)$, between 0.5 and 1.9. The dashed rectangle represents the field of view of IRS spectra in mapping mode.}

\label{fig.irac}
\end{figure}

The 8{\mm} filter includes the PAH emission from C-C stretching and C-H in plane bending modes at 7.6, 7.8 and 8.6{\mm} \citep{Leger1984,Tielens2008} while the 3.6{\mm} filter samples only the stellar continuum. To obtain an image of pure PAH emission we scaled down and subtracted the 3.6{\mm} image from the 8{\mm}. The subtraction operation can be summarised as

\begin{equation}
F_\nu(PAH) = F_\nu(8 \mu m) - \alpha F_\nu(3.6\mu m), 
\label{iracsub}
\end{equation}

\noindent where $F_\nu(PAH)$ is the flux density of non-stellar sources (mainly PAH emission), $F_\nu(8\mu m)$ and $F_\nu(3.6\mu m)$ are the original flux densities of the 8{\mm} and 3.6{\mm} images and $\alpha$ is the scaling factor for stellar flux at the 8{\mm} filter. Using the Starburst99 \citep{Leitherer1999} models \citet{Helou2004} demonstrated that the ratio of stellar flux between the 8{\mm} and 3.6{\mm} is almost independent of metallicity and star formation history, thus validating the use of a constant factor to remove the stellar component from the 8{\mm} image. Following their work we set $\alpha = 0.232$ as the scaling factor. The resulting images are shown in figure \ref{fig.irac} where we see interstellar emission in NGC~1386 extending up to distances of 1 kpc from the galactic centre, resembling a disk, and well defined spiral arms of NGC~7213.

\subsection{Spitzer IRS}

We used archival data from the IRS in mapping mode, which produces 3D spectra similar to an integral field unit (IFU) by taking several long-slit spectra in different positions. The original observing program ID is 3269 by Jack Gallimore (PI) and the observations were made in April 4, 2005. These data were first published by \citet{Wu2009} and have also appeared in a paper by \citet{Gallimore2010}, both of which have focused on integrating the spaxels (spatial picture elements that result from the virtual apertures used to extract the spectra) to produce spectroscopic data with very accurate aperture definitions. In this work we took a different approach as we try to understand the nature of MID-IR sources in different regions interior to and including the extended PAH emission.

The software {\sc cubism} \citep{Smith2007} was employed to build and analyse the data cube from the IRS spectra starting from the basic calibration data (BCD) set. Background levels were evaluated from off-source exposures of complementary positions along the slit, e.g. the second order sampled the background while the object was centred in the first order.

The wavelength coverage of this observation extends from 5.1{\mm} to 40.0{\mm} with a resolving power between 60 - 127 in multiple spectroscopic orders, which corresponds to the modules Short-Low (SL) from 5 to 14{\mm} and Long-Low (LL) from 14{\mm} to 40{\mm}. The plate scale of each setting is 1.8 pixel/\arcsec and 5.1 pixel/\arcsec for SL and LL respectively. The larger slit width and plate scale of LL sets the lower limit in spatial resolution for a full spectrum in 10.7\arcsec. 

\begin{figure}
\begin{center}
\includegraphics[width=\columnwidth]{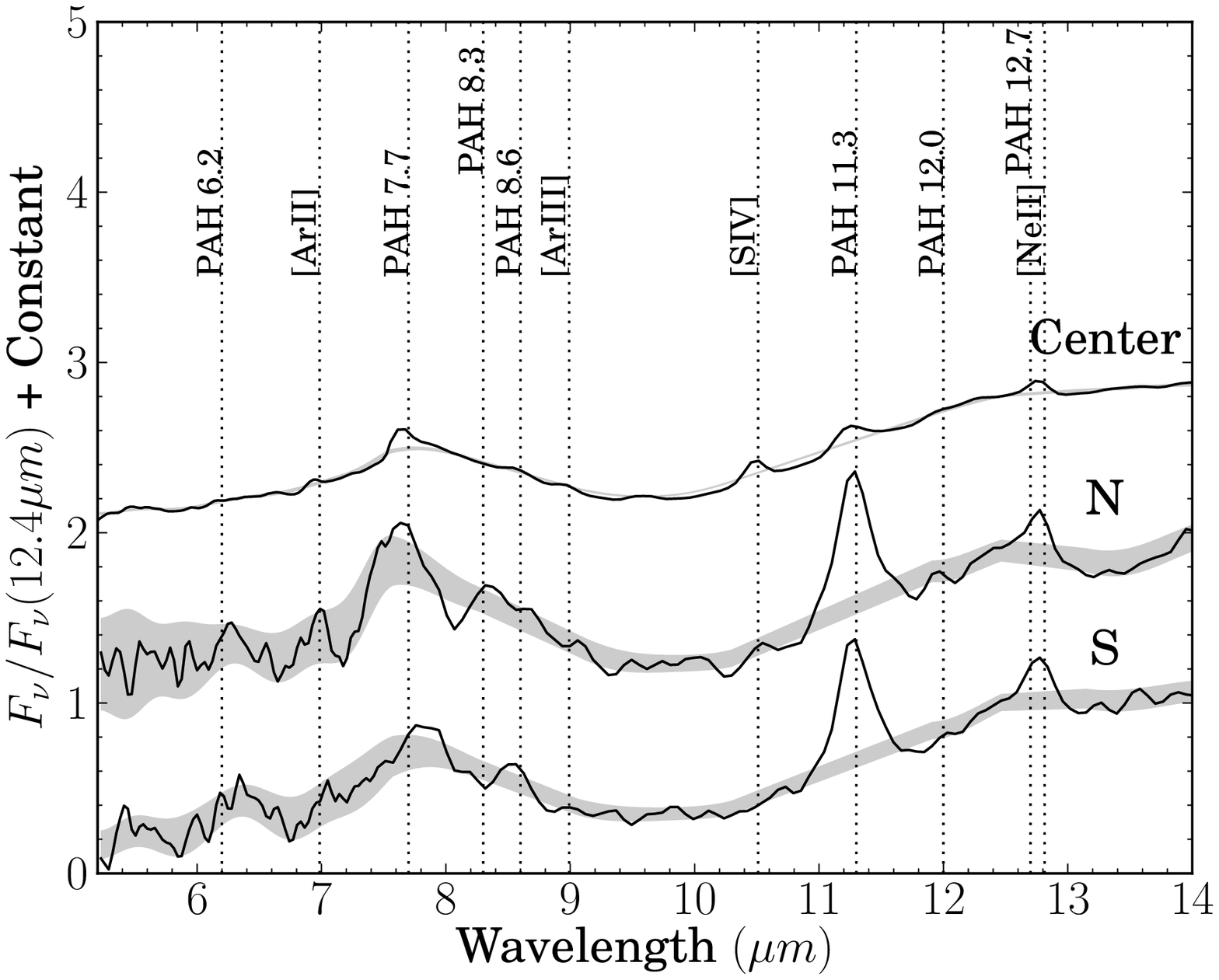}
\includegraphics[width=\columnwidth]{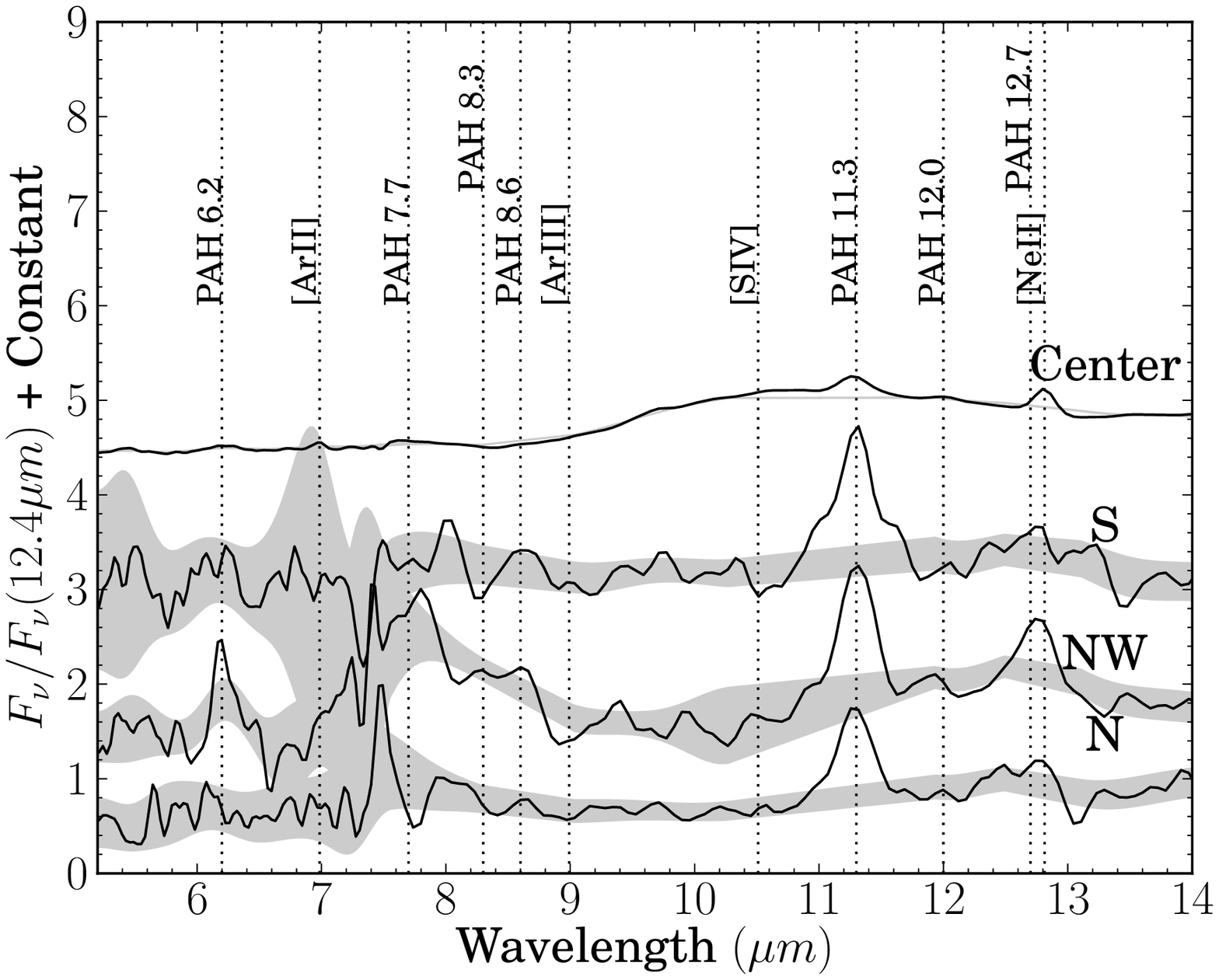}
\end{center}
\caption{ Spectra from IRS map for NGC~1386 (\textit{top frame}) and NGC~7213 (\textit{bottom frame}). PAH bands are detected in every spectra. Also the high ionisation line [S{\sc iv}] is present only in the central extraction of NGC 1386. The shaded areas represent the uncertainty evaluated by {\sc cubism} during the extraction.}
\label{fig.irsspec}
\end{figure}

Since we are interested in resolving the nucleus and the larger structure seen in the IRAC images we focused on the SL module, which allows extractions corresponding to a spatial resolution a factor of three higher than that of the LL module, but has the disadvantage of only sampling the spectrum from 5{\mm} to 14{\mm}. The apertures seen as white circles in figure \ref{fig.irac} were chosen for being representative of the PAH emitting regions of the host galaxy, or being its nucleus. It should be noted that the IRS mapping, which samples the region delimited by the dashed rectangles in figure \ref{fig.irac}, does not cover the entire field of view of the IRAC images. Therefore brighter regions in figure~\ref{fig.irac} were not chosen because they fall outside the spectroscopically sampled area.
We used circular apertures with a radius of 3.6\arcsec which correspond to $\sim$210pc for the first galaxy and $\sim$420pc for the last one. Spectra from these extractions are shown in figure \ref{fig.irsspec}.



\subsection{T-ReCS}

In an effort to probe in more detail the nuclear region of the galaxies we obtained observations with Gemini South's T-ReCS, with an angular resolution of 0.36\arcsec. The data used here correspond to the 10{\mm} low resolution mode of T-ReCS, with a resolving power $R \sim 100$ and wavelength range between 8 and 13{\mm}. The slit has a projected width of $0.35\arcsec$ which correspond to 21 pc and 41 pc for NGC~1386 and NGC~7213 respectively.

Removal of atmospheric emission was achieved via the standard chopping and nodding technique, with chop throws of 15\arcsec. Individual frames have an integration time of 43.1 ms and for each of the chop cycles in one save set 3 frames are co-added. The total on-source time is 981~s and 1248~s for NGC~1386 and NGC~7213 respectively.

All the reduction process was performed with Gemini's {\sc midir} package for {\sc iraf} \citep{Tody1986,Tody1993}. Most of the work, like chop-nod subtractions, is done automatically with only the aperture definition, tracing, telluric correction and wavelength calibration done interactively. Wavelength calibration was based on the identification of telluric emission lines in the sky spectra. A fourth-order Chebyshev polynomial was fitted to five strong features in the off-source spectra resulting in a typical RMS below 70 \AA. These same five features were then automatically re-identified in different positions along the spatial axis to produce an interactive fit of wavelength versus 2D position. In order to guarantee less than an hour of interval between science and telluric calibration exposures, two different standard stars were observed for each galaxy.

\begin{figure}
\includegraphics[width=\columnwidth]{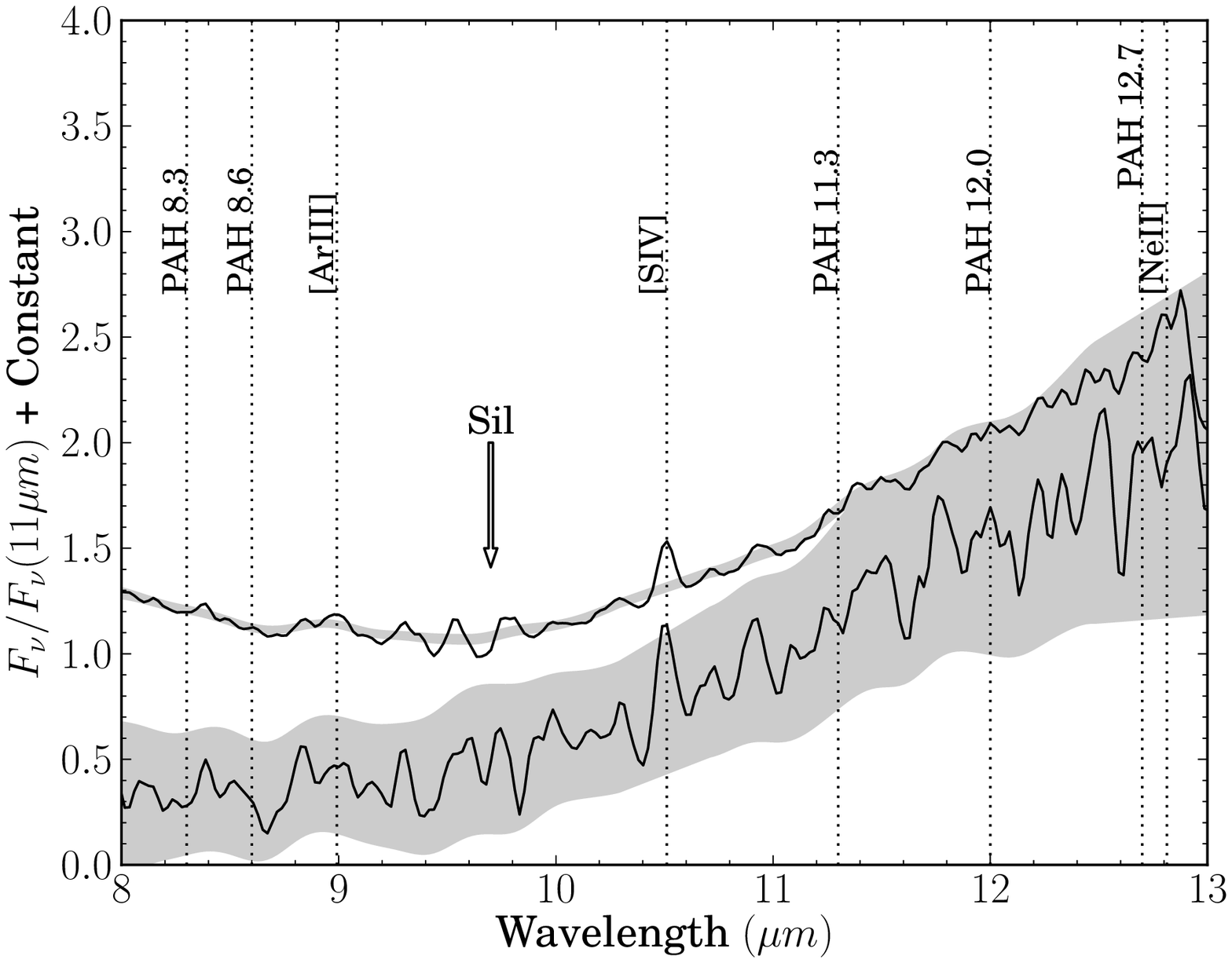}
\includegraphics[width=\columnwidth]{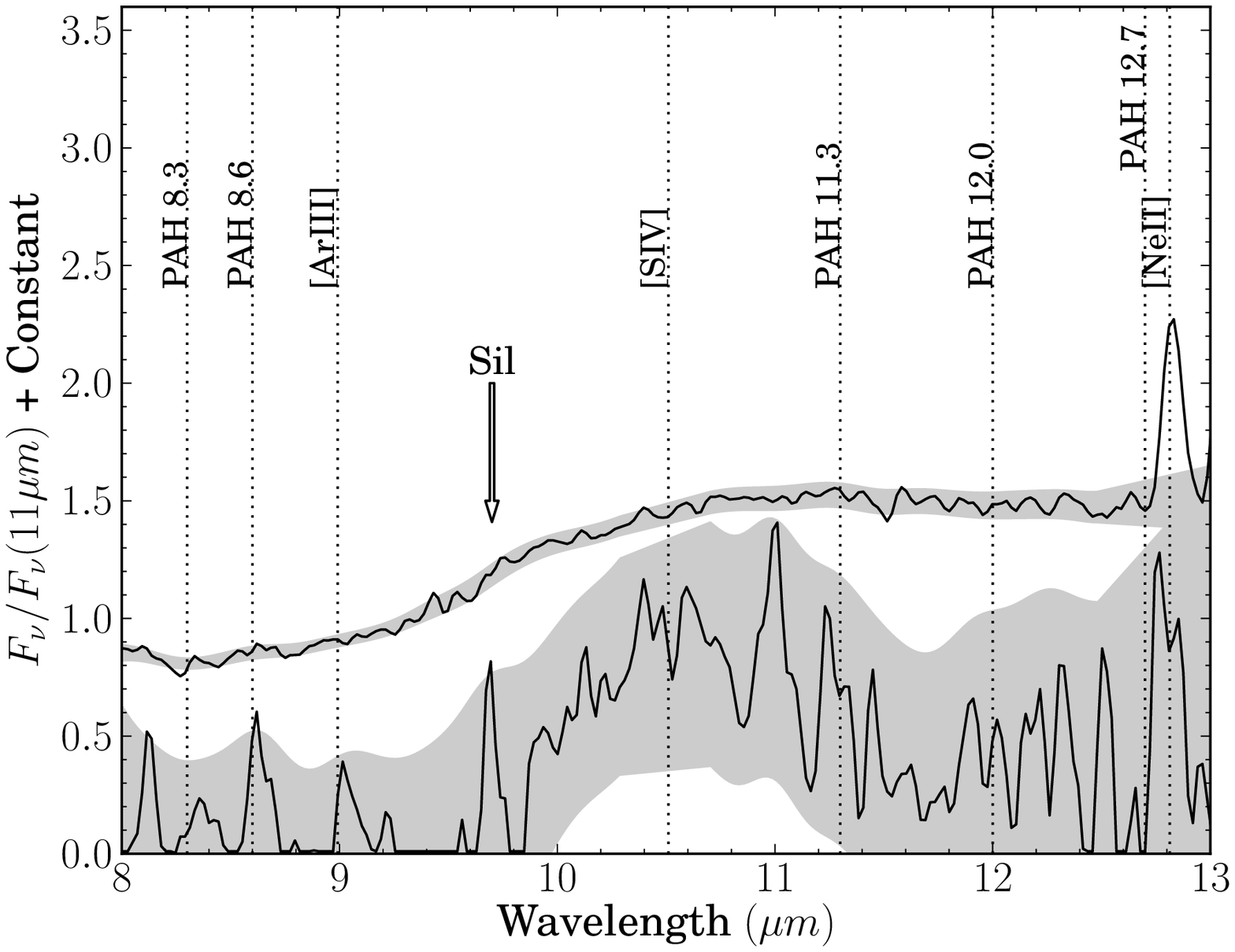}
\caption{Nuclear (\textit{top}) and extended (\textit{bottom}) spectra of NGC~1386 (\textit{top frame}) and NGC~7213 (\textit{bottom frame}) obtained with T-ReCS. Shaded regions are limited by the Poisson noise convolved by a Gaussian with $\sigma = 5$ px. The arrow marks the centre of the silicate absorption feature, which extends through almost the entire spectrum.}
\label{fig.trecsspec}
\end{figure}

Since the spatial profile of these galaxies' spectra is very close to that of a point source, partly due to the bright AGN, we expect a large contamination of the extended emission by the nuclear source. In order to obtain a pure extended spectrum two extractions were performed: one having the width of the stellar FWHM and the other encompassing all the light available from the source, meaning almost 4 times the stellar FWHM. The extended emission spectrum was obtained by subtracting 1.314 times the central spectrum from the largest possible aperture. This corresponds to the infinite aperture correction for a FWHM extraction of a Gaussian profile. In figure \ref{fig.trecsspec} we show the nuclear and outer spectra. Shaded areas in figure \ref{fig.trecsspec} represent the Poisson noise, estimated directly from the instrumental analog to digital units, smoothed by a Gaussian with $\sigma = 5$px. The main features of the nuclear spectrum of NGC~1386 are the silicate absorption at 9.7{\mm} and the [S{\sc iv}] line, while the host galaxy spectrum shows only an increase in flux towards longer wavelengths due to the interstellar dust. The extended emission at 11.8{\mm} identified by \citet{Reunanen2010} can not be confirmed by our data. The low ionisation characteristic of NGC~7213's nuclear spectra, already identified in the optical spectra \citep{Filippenko1984}, is revealed in the MIR by the prominent [Ne{\sc ii}] at 12.7{\mm} and complete absence of the [S{\sc iv}] line. In the nuclear spectrum we also see the silicate emission which causes a rise in the continuum above 9{\mm}. The host galaxy spectrum resulted of too low S/N ratio to allow any conclusive analysis.


\section{Off-nuclear Star Formation}

The main goal associated with the Spitzer data is to study the star formation rate (SFR) in selected regions of the galaxies. To assess it two different indicators have been used: a) the 12.8{\mm} [Ne{\sc ii}] ionic line from the IRS mapping (hereafter [Ne{\sc ii}] method); b) the photometric flux of the stellar subtracted 8{\mm} image. They will be discussed in detail in section \ref{results.sfr}.

It is worth mentioning that the studied regions were selected based solely on the intensity of the 8{\mm} image. An aperture of 3.6\arcsec in diameter was chosen leading to a spatial sampling of twice the slit width per element, although the IRAC images would allow smaller apertures. This angular diameter corresponds to 416 pc and 208 pc at the distances of NGC~7213 and NGC~1386 respectively. The chosen locations are indicated in figure \ref{fig.irac}

\subsection{IRAC images}

Continuum subtracted 8{\mm} images (hereafter PAH8, figure \ref{fig.irac}) were obtained following equation \ref{iracsub}. In the extended regions they reveal practically only the interstellar emission, which in this case consists of PAH bands and possibly a small contribution from hot dust. It should be noted that equation \ref{iracsub} assumes a stellar population, therefore the AGN's power law continuum will inevitably remain after the subtraction that leads to the PAH8 image. A similar scaling could be applied to remove the non-thermal nuclear component only if we had continuum samples closer to the 8{\mm} features.

\begin{figure}
\begin{center}
\includegraphics[width=\columnwidth]{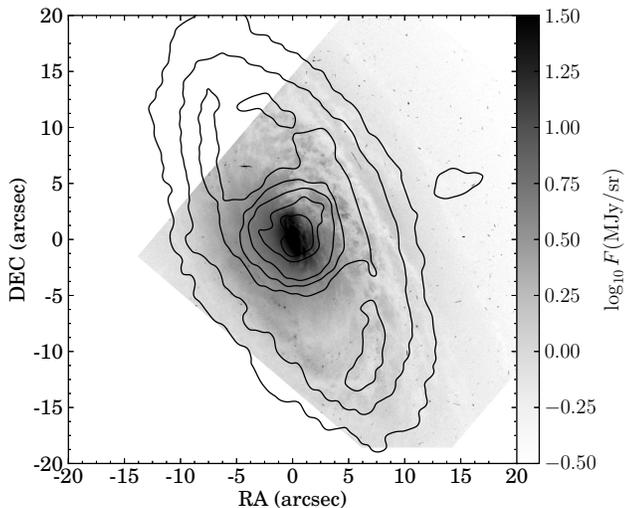}
\end{center}
\caption{Contours from figure \ref{fig.irac} superimposed to HST's image with the F606W filter. Dust lanes seen as highly absorbed regions in the optical image are mostly followed by intense PAH emission.}
\label{fig.hst}
\end{figure}

In both galaxies the emission is characterised by an intense flux at the nucleus (mainly due to the power law continuum) surrounded by larger structures along the spiral arms, which in NGC~1386 resembles an elongated ring. An ellipse that fits the peak intensity of this ring in the radial direction has semi-axes of 6.5\arcsec and 14.5\arcsec with the major axis having a PA of 25\degr. Not considering the intrinsic width of the ring we obtain an inclination of 65\degr which is very close to the value of 71{\degr} found by \citet{Rossa2000} for the galaxy's disk. The semi-major axis correspond to nearly 750 pc from the galaxy nucleus to the centre of the ring. The distribution of the PAH emission can be related to the various dust lanes seen in optical images of the Hubble Space Telescope (HST). In figure \ref{fig.hst} we show the contours from figure \ref{fig.irac} superimposed to the F606W image of HST's Wide-Field Planetary Camera 2 (WFPC2).

In NGC~7213 we see that the interstellar emission follows the spiral arms reaching distances of up to 10 kpc from the centre. Additionally we see several H{\sc ii} nodes distributed along the bright spiral arm (see figure \ref{fig.irac}), with the one in the NW extraction being the most luminous.

Fluxes for the IRAC images were obtained by summing the flux of each pixel inside the defined apertures. No corrections were made to account for partial pixels, since a 3.6{\arcsec} aperture correspond to more than 110 pixels in the IRAC images, and variations in the number of pixels in different apertures are less than 5\%. The derived luminosities for each of the apertures are presented in table \ref{tab.apsum}. The centre apertures have almost 15 times the flux of the apertures at the host galaxy, partially because radiation from the AGN was not removed by subtraction of the stellar component.

\begin{table}
\centering
\caption{Luminosities measured in Spitzer's images and spectra}
\begin{tabular}{l c c}
Region & $L({\rm PAH8}) \times 10^{-42}$ & $L(\mbox{[Ne {\sc ii}]}) \times 10^{-38}$ \\ 
& (erg s$^{-1}$) & (erg s$^{-1}$) \\
\hline
NGC~1386 & \\
\hspace{12pt} Centre & 35.38 & 12.70\\ 
\hspace{12pt} N & 2.55 & 1.96\\ 
\hspace{12pt} S & 2.34 & 2.06\\ 
\hline
NGC~7213 & \\
\hspace{12pt} Centre & 78.86 & 78.44\\
\hspace{12pt} NW & 5.64 & 5.82\\
\hspace{12pt} E & 4.11 & 3.95\\
\hspace{12pt} S & 3.30 & 3.76\\
\hline
\end{tabular}
\label{tab.apsum}
\end{table}


\subsection{IRS spectra}

The one dimensional spectra shown in figure \ref{fig.irsspec} (extracted in the same locations used to obtain the photometric fluxes) were used to measure the fluxes and equivalent widths (EW) of PAH bands and ionic lines through the PAHFIT code \citep{Smith2007a}. By fitting numerous spectral features of both atomic and molecular origin, along with dust grains emission and absorption and a stellar continuum, this code is capable of producing flux measurements of several spectral features without recurring to a simplistic linear continuum fit. It should be noted that PAHFIT's primary motivation for fitting atomic lines is to reach a better fit of the PAH features, thus these measurements might not be as precise as those for the latter. Table \ref{tab.eqw-irs} shows the EW of some prominent features in the IRS spectrum, and table \ref{tab.apsum} shows the luminosity of the 12.8{\mm} [Ne{\sc ii}] line. The high ionisation 10.5{\mm} [S{\sc iv}] line is only detected in the nucleus of NGC~1386. The PAH features are notably more pronounced in the outer extractions than in the nucleus in both galaxies.

\begin{table*}
\centering
\caption{EW of selected features in the IRS spectra in units of{\mm}.}
\begin{tabular}{l c ccc cccc}
\hline
&& \multicolumn{3}{c}{NGC 1386} & \multicolumn{3}{c}{NGC 7213} \\ \cline{3-5} \cline{6-9}
Feature & $\lambda$ ({\mm}) & $W_\lambda$ Centre & $W_\lambda$ N & $W_\lambda$  S & $W_\lambda$ Centre & $W_\lambda$ E & $W_\lambda$ NW & $W_\lambda$ S \\
\hline
[S{\sc iv}] & 10.5 & 0.08 & -- & -- & -- & -- & -- & --\\ 

[Ne{\sc ii}] & 12.8 & 0.02 & 0.06 & 0.06 & 0.08 & 0.10 & 0.25 & 0.28\\

PAH & 6.2 & -- & 0.34 & 0.23 & 0.04 & 0.09 & 6.15 & 0.04\\
PAH & 7.7 & 0.36 & 1.56 & 1.00 & 0.13 & 0.35 & 7.29 & 0.51\\
PAH & 8.6 & -- & 0.94 & 0.51 & -- & 0.25 & 2.38 & 1.07\\
PAH & 11.3 & 0.12 & 1.00 & 0.75 & -- & 0.94 & 2.22 & 2.06\\
\hline
\end{tabular}
\label{tab.eqw-irs}
\end{table*}

The NW aperture in NGC~7213 has notably the largest EW of PAH bands, which suggests intense star formation, or a very young stellar population.  \citet{Genzel1998} shows that the ratio  7.7{\mm} PAH  feature/7.7{\mm} continuum decreases from starburst galaxies to AGN. Similar behaviour is observed for the PAH bands at EW 7.7{\mm} and EW 11.3{\mm} for a sample of starburst galaxies and AGNs. Galaxies powered by star formation have a larger EW than Sy1 and Sy2 \citep[see figure~9 in ][]{Sales2010}.

The intense star formation in the NW region of this galaxy is  confirmed by the optical study of \citet{StorchiBergmann1996}. Although the apertures in that study are not directly comparable to the ones in this paper, the H$\beta$ line in the extraction over NW region has a high EW, which agrees with our conclusion.

\subsection{Star Formation}
\label{results.sfr}

In this section we investigate the star formation in the host galaxies using the PAH8 images and the ionic line 12.8{\mm} [Ne{\sc ii}]. The photometric estimation of SFR from the PAH8 images depends on an accurate subtraction of the underlying continuum sampled by the 3.6{\mm} filter, which in turn is based on an assumption of relative fluxes between these two filters. Our approach here was to use the already published results of \citet{Helou2004}, designed to remove the stellar component. However the power law continuum of the AGN behaves very differently, and the uncertainties in the exponent would be comparable or greater than the measured values for SFR. In the case of the [Ne{\sc ii}] line, one can never be sure that the AGN is not contributing to the gas' excitation without recurring to additional spectral features, thus rendering this indicator susceptible to sources of radiation completely unrelated with young stars. For these reasons we refrain from any estimates of SFR in the nucleus of these galaxies and focus only on the extended regions.

\begin{figure}
\begin{center}
\includegraphics[width=\columnwidth]{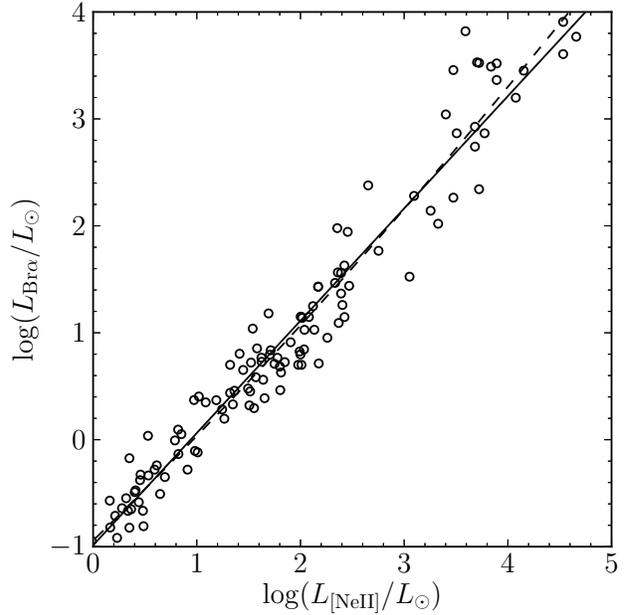}
\end{center}
\caption{Correlation between the luminosities of 12.8{\mm} [Ne{\sc ii}] and Br$\alpha$ \citep{Giveon2002,Willner2002}. The solid line represents a linear fit while the dashed line is a second order polynomial.}
\label{fig.nebra}
\end{figure}

\citet{Ho2007} demonstrated a correlation between the sum of luminosities of 12.8{\mm} [Ne{\sc ii}] and 15.7{\mm} [Ne{\sc iii}] lines and the SFR measured from Hydrogen recombination lines, Br $\alpha$ in this particular case. This of course warrants the use of these fine structure features as indicators of star formation rate. In figure \ref{fig.nebra} we reproduce the data from \citet{Giveon2002,Willner2002} for H{\sc ii} regions in the Galaxy, Magellanic Clouds and M33, showing the empirical correlation between $L_{[{\rm Ne II}]}$ and $L_{{\rm Br}\alpha}$. It is clear that there is a fair correlation between these two indicators, particularly for lower luminosities. At the high end of the graph there is much more scattering in the [Ne{\sc ii}] luminosity, reaching more than an order of magnitude between the last three points. Since the luminosities in our sample of star forming regions are high we should expect a somewhat less reliable assessment of Br$\alpha$ luminosity. In the high luminosity regime only the sum of the above mentioned Ne features retain its linear correlation with Br$\alpha$ while the 12.8{\mm} line strays away from linearity towards lower values. Nevertheless there simply are not enough empirical points in figure \ref{fig.nebra} to justify a second order fit, thus we employ a linear correlation to yield the corresponding Br$\alpha$ luminosity


\begin{equation}
\log \frac{L_{{\rm Br}\alpha}}{L_\odot} = 1.05 \times \log \frac{L_{[{\rm Ne II}]}}{L_\odot} - 0.99.
\label{sqrfit} 
\end{equation}




To derive the SFR one can choose from the vast number of available relations between the Lyman continuum and H emission. Here we used the SFR estimate based on the H$\alpha$ luminosity \citep[see equation 2 of][]{kennicutt1998} for a Salpeter IMF with masses in the range 0.1$M _\odot$ - 100$M_\odot$ and solar abundances. In order to obtain a SFR in terms of $L_{\rm Br \alpha}$ we used the intrinsic intensity ratio $j_{\rm H \alpha}/j_{\rm Br \alpha} = 35.75$, assuming case B recombination for an electronic temperature $T_e = 10^4$\,K and density $n_e = 10^2$ \citep{Brocklehurst1971}.

\begin{equation}
SFR ({\rm M_\odot yr^{-1}}) = 2.82 \times 10^{-40} L_{\rm Br \alpha} ({\rm erg\,\,s^{-1}}).
\label{sfr}
\end{equation}



Using equations \ref{sqrfit} and \ref{sfr} together with the luminosity of the 12.8{\mm} [Ne{\sc ii}] line we can estimate the SFR. Since the distances to the two galaxies are vastly different we base the following discussion in the SFR divided by the projected area of the extraction, or the superficial density of star formation rate ($\Sigma_{\rm SFR}$). The obtained values for $\Sigma_{\rm SFR}$ and the estimated luminosity of Br$\alpha$ are presented in table \ref{tab.sfr}. The errors in $\Sigma_{\rm SFR}$ include the RMS of the fit of equation \ref{sqrfit}. Our estimates are comparable to those found by \citet{Kennicutt1998a} for normal spiral galaxies.


\begin{table*}
\caption{SFR surface densities and estimated $L_{Br\alpha}$}
\begin{tabular}{lccc}
\hline
Region & $L_{{\rm Br}\alpha} \times 10^{-38} $ & $\Sigma_{\rm SFR} ({\rm PAH8}) \times 10^{2}$ & $\Sigma_{\rm SFR} (\mbox{[Ne{\sc ii}]}) \times 10^2$\\
 
& (erg s$^{-1}$) & (${\rm M}_\odot\,{\rm yr}^{-1}\,{\rm kpc}^{-2}$) & (${\rm M_\odot \,yr^{-1}\,kpc^{-2}}$)\\
\hline
NGC~1386\\
\hspace{12pt}N & 0.61 & $111 \pm 28$ & $7.2 \pm 2.0$\\
\hspace{12pt}S & 0.64 & $102 \pm 26$ & $7.6 \pm 2.1$\\
\hline
NGC~7213\\
\hspace{12pt}NW & 2.38 & $62 \pm 16$ & $5.8 \pm 1.6$ \\
\hspace{12pt}E & 1.44 & $45 \pm 11$ & $3.8 \pm 1.0$\\
\hspace{12pt}S & 1.36 & $36 \pm 9$ & $3.6 \pm 1.0$\\
\hline
\end{tabular}
\label{tab.sfr}
\end{table*}


To test the results derived from the ionic emission we compared those values with the completely independent SFRs derived from the photometric PAH fluxes from IRAC (table \ref{tab.sfr}). Following the work of \citet{Wu2005} we used the relation 

\begin{equation}
SFR ({\rm M_\odot\,yr}^{-1}) = \frac{\nu L_\nu[8 \mu m]}{1.57\times10^9 L_\odot}.
\label{sfrpah}
\end{equation}

\begin{figure}
\begin{center}
\includegraphics[width=\columnwidth]{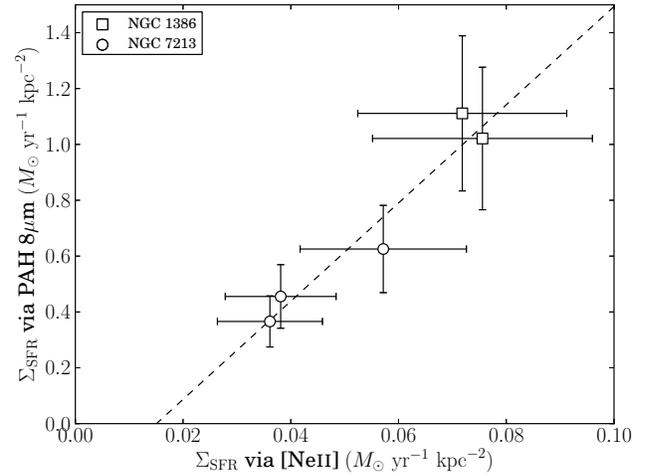}
\end{center}
\caption{Comparison between the SFR surface density for the three regions of NGC~7213 and two regions of NGC~1386. It is clear that the results based on the photometric flux of the PAH bands at 8{\mm} consistently overestimate $\Sigma_{\rm SFR}$.}
\label{fig.nephot}
\end{figure}

Figure \ref{fig.nephot} shows the correlation between the SFR surface density $\Sigma_{\rm SFR}$ derived from each approach. A linear fit to the data has an inclination of $18\pm8$ and intercepts $\Sigma_{\rm SFR} ({\rm PAH})$ at $-0.3\pm0.5$. In other words there is a factor of more than one order of magnitude between the two estimates. Errors in the subtraction of stellar emission and other unrelated interstellar sources could be at least partially responsible for the photometric overestimate. This effect has already been reported by \cite{Calzetti2007}, where the authors argue that these PAH bands are also susceptible to excitation from photons of older stellar populations. Moreover the spread in the high luminosity end of figure \ref{fig.nebra} can easily explain this discrepancy, even without invoking any errors in the photometry.

Despite the disagreement between both SFR indicators the general picture is internally preserved, with the H{\sc ii} regions of NGC~1386 having a higher $\Sigma_{\rm SFR}$ than their counterparts in NGC~7213. There is no particular reason why the star formation in any circumnuclear ring should be higher than any random H{\sc II} region in a spiral galaxy, so these two results alone cannot be extrapolated to a more general conclusion. Furthermore the galactocentric distances of the extractions set the extended spectra of NGC~1386 inside the region that is thought to suffer major influence of the AGN ($<$ 1 kpc), an interpretation which is further supported by the close agreement between the SFRs of these two symmetrically located extractions. The same reasoning does not hold for NGC~7213 because the three extractions are beyond 2 kpc from the nucleus.

\section{Nuclear spectra: a dusty torus signature?}

\label{results.nuclear}

The high spatial resolution of the Gemini telescope allows us to probe regions within tens of parsecs from the active nucleus. In such small scales a dramatic change in some spectral features is expected as the structures associated with the AGN begin to dominate the light. The silicate absorption at 9.7{\mm} is one of such features since it is expected to be linked to the dusty torus surrounding the accretion disk required by the unified AGN model \citep{Nenkova2002,Nenkova2008,Nenkova2008a,Sales2011}. In outer regions, where stars dominate the light, the silicate should be less prominent if detectable at all. 

In figure \ref{fig.trecsspec} we present the nuclear and extended spectra for both galaxies. Although the presence of PAH bands is clearly visible in all the spectra from IRS, the same bands are completely absent in the higher spatial resolution spectra of T-ReCS. We can thus safely set a lower limit for the galactocentric distances below which PAH cannot be found as 21 pc for NGC~1386 and 42 pc for NGC~7213. This observation is compatible with the already observed anti-correlation between the AGN radiation and PAH emission \citep{O'Dowd2009,Treyer2010,LaMassa2012}.

A shallow but visible depression centred in the silicate band is visible in the nuclear spectrum of NGC~1386 while the outer one is almost flat before 10{\mm}.  Following the method described by \citet{spoon2007} we measured the strength of this silicate absorption ($S_{\rm sil}$) by adjusting a linear continuum to the median fluxes at 1{\mm} windows centred in 8.2{\mm} and 12.4{\mm} \citep{Mason2006} and taking the ratio between observed and continuum fluxes at 9.7{\mm}. In symbols 

\begin{equation}
S_{\rm sil} = \log \frac{f_{\rm obs}(9.7\,\mu m)}{f_{\rm cont}(9.7\,\mu m)}.
\label{ssil}
\end{equation}

\begin{figure}
\begin{minipage}[b]{.48\columnwidth}
\includegraphics[width=\columnwidth]{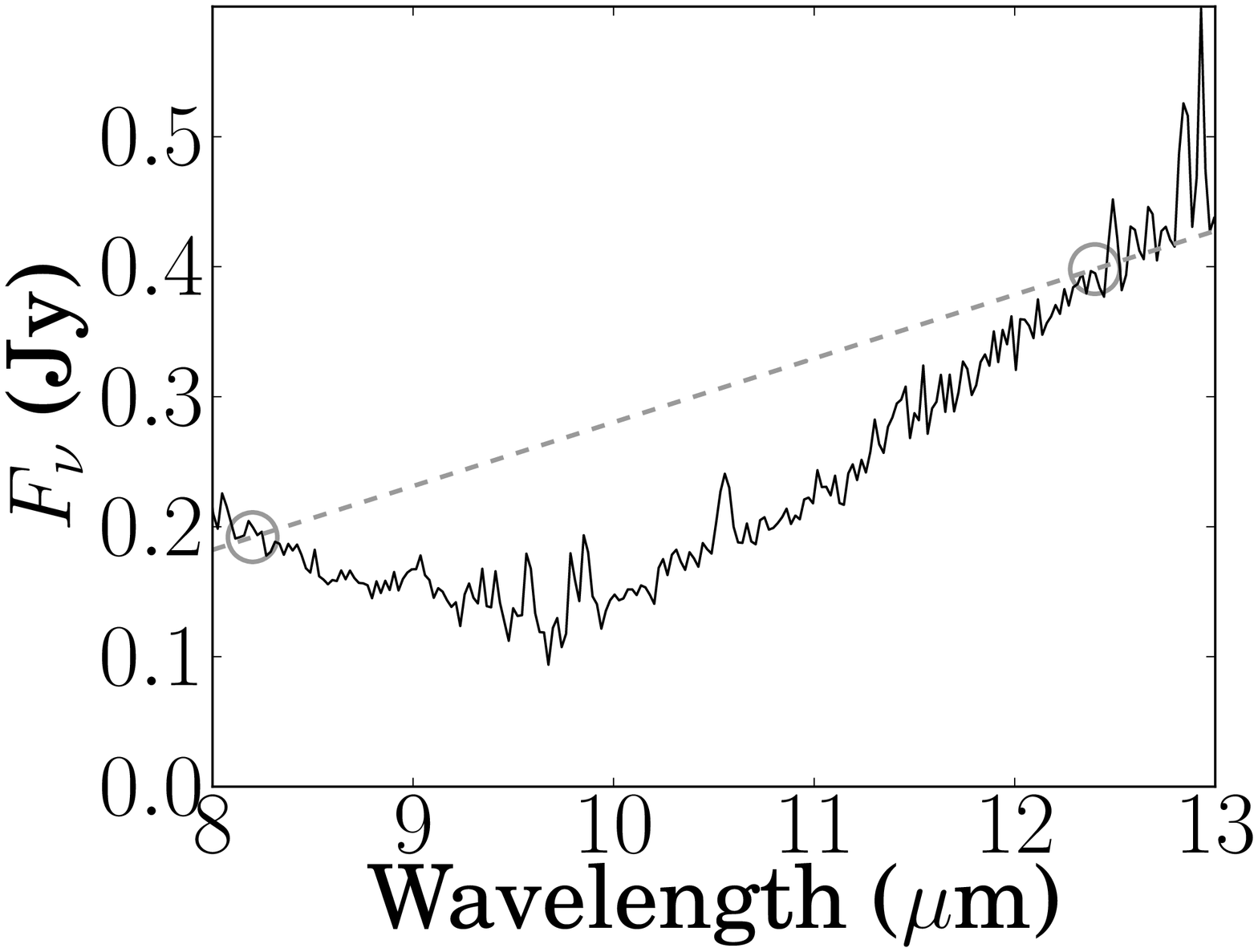}
\end{minipage}
\begin{minipage}[b]{.48\columnwidth}
\includegraphics[width=\columnwidth]{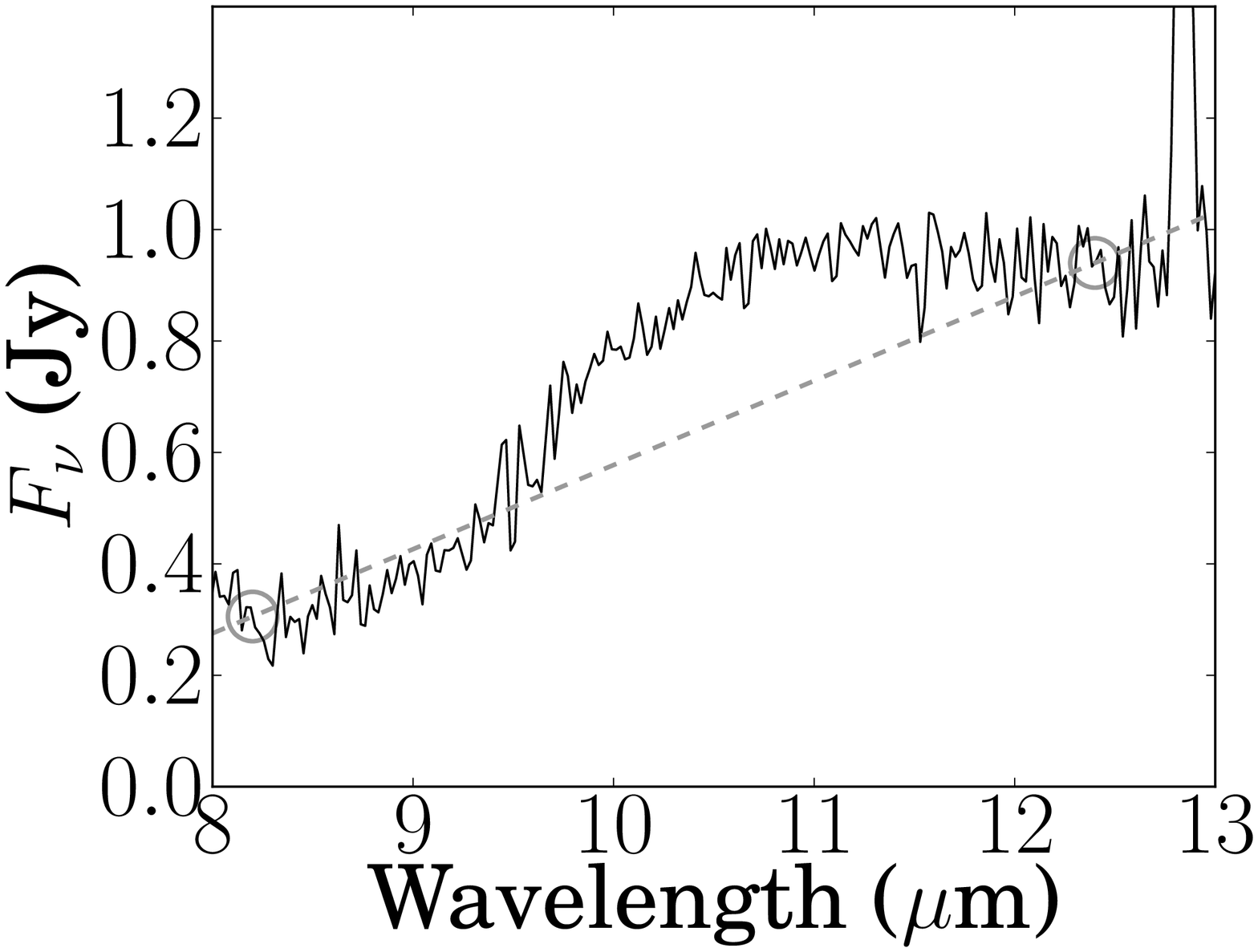}
\end{minipage}
\caption{Continuum fits to evaluate the silicate strength $S_{\rm{sil}}$ (equation \ref{ssil}) for NGC 1386 (\textit{left}) and NGC 7213 (\textit{right}). The dashed grey line represents the inferred continuum and grey circles are at the centre of the two sample regions used for the fit.}
\label{fig.ssil}
\end{figure}

\noindent The inferred continuum for the central extraction of both NGC~1386 and NGC~7213 is plotted in figure \ref{fig.ssil}. We obtain $S_{\rm sil} = -0.69^{+0.19}_{-0.23}$ and $S_{\rm sil} = -0.42^{+0.30}_{-0.44}$ for the central extraction and outer region of NGC~1386 respectively. These values agree with the published results obtained by \citet{Wu2009} for an aperture of $20.4 \arcsec \times 15.3\arcsec$. The only detectable emission feature in the central extraction is the 10.5{\mm} [S{\sc iv}] line.

Approximately at 9{\mm}  in the nuclear spectrum of NGC~7213 we begin to see a rise in the continuum level caused by the emission from silicate grains with an estimated $S_{\rm sil} = 0.15^{+0.14}_{-0.16}$. The low ionisation line of [Ne{\sc ii}] at 12.8{\mm} is present in the nuclear spectrum but the high ionisation [S{\sc iv}] is absent. Since the published optical spectrum \citep{Filippenko1984} of this galaxy shows lines of higher ionisation potential such as [O{\sc iii}] and [Ne{\sc v}], the absence of the 10.5{\mm} feature must be linked to the density of the interstellar medium rather than to the ionisation parameter. Indeed the transition which leads to 12.8{\mm} [Ne{\sc ii }] has a critical density more than one order of magnitude higher than that of the 10.5{\mm} [S{\sc iv}]. We therefore conclude that the density in the narrow line region of NGC~7213 is at least $5 \times 10^4 \,\, {\rm cm}^{-3}$, which agrees with the findings of \citet{Filippenko1984}.

\subsection{Fitting of torus models}

{\sc clumpy}\footnote{The models are available on-line at the following address: http://www.pa.uky.edu/clumpy/} torus models \citep{Nenkova2002,Nenkova2008} were used to study the continuum of the nuclear spectrum. Following \citet{Sales2011} the emission lines and the telluric band region were masked and the resulting clear nuclear spectrum was compared to the {\sc clumpy} theoretical SEDs. These SEDs were generated assuming that the torus is formed by dusty clumps, constrained as follow: (\textit{i}) the number of clumps, $N_{0}$, along to the torus equatorial radius; (\textit{ii}) the visual optical depth of each clump, $\tau_{V}$;  (\textit{iii}) the radial extension of the clumpy distribution, $Y=R_0/R_d$, where $R_0$ and $R_d$ are the outer and inner torus radii, respectively; (\textit{iv}) the radial distribution of clouds as described by a power law $\propto r^{-q}$; (\textit{v}) the torus angular width, constrained by a Gaussian angular distribution described by a half-width $\sigma$; (\textit{vi}) the observer's viewing angle $i$.

The best fitting parameters for the observed nuclear spectrum were obtained via the BayesCLUMPY inference tool \citep{AsensioRamos2009}. By using a Markov Chain Monte Carlo method the code investigates a parameter space defined by the first 13 eigenvectors of the principal component analysis of the model grid with more than $10^6$ individual SEDs. An artificial neural network then assesses the marginal posterior distribution for each of the parameters taking into account the observations. The only \textit{a priori} information given to BayesCLUMPY is a limitation in the line of sight angle, which was greater than 45\degr for the Sy2 and less than 45\degr for the Sy1. The stability of the solution was confirmed by consecutive runs of the algorithm.

\begin{table}
\centering
\caption{{\sc clumpy} best fit parameters.} 
\label{tab.clumpy} 
\begin{small}
\begin{tabular}{l cc}
\hline \hline
\noalign{\smallskip}
Parameter & \multicolumn{2}{c}{Value with 68\% Confidence}\\ \cline{2-3}
& NGC~1386 & NGC~7213 \\
\hline
$\sigma$ & $68_{-2}^{+1}$ & $21_{-3}^{+4}$\\
Y & $18_{-3}^{+53}$ & $79_{-13}^{+10}$\\
N$_0$ & $14.6_{-0.5}^{+0.2}$ & $3.0_{-0.7}^{+0.7}$\\
q & $2.5_{-0.1}^{+0.1}$ & $0.3_{-0.1}^{+0.2}$\\
$\tau_V$ & $65_{-3}^{+3}$ & $44_{-12}^{+16}$\\
i & $81_{-8}^{+6}$ & $21_{-12}^{+9}$\\
\noalign{\smallskip}
\hline
\end{tabular}
\end{small}
\end{table}


It is important to keep in mind that fitting {\sc clumpy} torus models to the spectra is an intrinsically degenerate problem, as changes in a given set of parameters can produce the same observable effect as different changes in another set. For instance, the number of clouds and the optical depth of a singular cloud can both account for the reddening of the spectra. The degeneracy is especially notable when dealing with narrow wavelength ranges such as ground-based MIR observations. On the other hand, this spectral window encompasses one of the major spectral features attributed to the torus, the 9.7{\mm} silicate feature, which is very relevant in anchoring the models.

The results obtained are in agreement with the general picture of the unified model, with the inclination tending towards low values for the Sy1 NGC~7213 and high values for the Sy2 NGC~1386. It is interesting to note that the inclination found for NGC~1386 puts the torus in the same plane of the galaxy's disk. In this particular case, where dust lanes are a major feature of the galaxy's structure, this coincidence could be evidence that the host galaxy is influencing the parameter estimates for the torus, especially when the fitting relies so heavily upon the silicate feature. This would agree with recent works that attribute the bulk of silicate absorption, in sources which show intense obscuration, to the host galaxy rather than the torus \citep{Goulding2012,Gonzalez-Martin2013}. Moreover there is evidence of a higher probability for Sy1's to live in face-on spiral galaxies \citep{Keel1980,Simcoe1997,Lagos2011}, although there are studies which show no alignment between the plane of the torus and that of the galaxy disk \citep{Schmitt1997,Kinney2000}.

Based on the best fit parameters summarised in table \ref{tab.clumpy} we can deduce the number of clouds in the viewer's line of sight $N_{\rm LOS}$. Adopting a standard dust-to-gas ratio the optical depth $\tau_V$ of a single cloud can be converted into the hydrogen column density for that cloud. Finally by combining these two results we get the hydrogen column density in the line of sight. The assumed distribution of clouds leads to the equation

\begin{equation}
N_{\rm H} = 1.5\times10^{21} \tau_V N\exp \left( -\frac{(90-i)^2}{\sigma^2}\right),
\label{nlos}
\end{equation}

\noindent therefore $N_{\rm H} = 1.4^{+0.1}_{-0.2} \times 10^{24}\,\,{\rm cm}^{-2}$ for NGC~1386 and $N_{\rm H} \sim  10^{18}\,\,{\rm cm}^{-2}$ for NGC~7213, which basically means that there are no clouds obscuring the latter. The results for both galaxies agree with the unified model for AGNs, in the sense that there is a clear line of sight to the Sy1 nucleus of NGC~7213 and an obscured one to the Sy2 in NGC~1386.

We emphasise that our results regarding the torus properties are based on the assumption that the main difference between the two types of Seyfert galaxies arise from the torus inclination. This hypothesis, however, has recently been challenged by \citet{RamosAlmeida2011a} with a similar analysis of a sample of 7 galaxies, where the authors conclude that the parameters $N_0$, $\sigma$ and $\tau_V$ are far more distinct than $i$, and therefore AGNs of types 1 and 2 have intrinsically different tori.

\section{Summary and conclusions}

In this work we present new spectral data taken with T-ReCS at the Gemini South Observatory for the Compton thick Sy2 galaxy NGC~1386 and the Sy1 galaxy NGC~7213. Additionally we also analyse and discuss archival data from the Spitzer Space Telescope both from imaging and spectral observations. Our conclusions are as follows:

\begin{enumerate}
\item PAH emission was not detected in the nuclear spectra of either galaxy. This is consistent with the view that PAH molecules are destroyed by the radiation field of the AGN. The strength of the later is confirmed in this paper by the presence of the S$^{+3}$ ion in the case of NGC~1386, and assumed from published optical studies for NGC~7213.
\item Using an empirical relation between the luminosity of the lines 12.8{\mm} [Ne{\sc ii}] and Br$\alpha$, and a well established relation between the Lyman continuum and star formation we estimate SFRs for two regions in NGC~1386 and three regions in NGC~7213. We interpret the close agreement between the first two as evidence of a common relation to the AGN. SFR indicators based on 8{\mm} photometry are 10 to 15 times higher than similar estimates based on the ionic line 12.8{\mm} [Ne{\sc ii}]. This effect could be linked to excitation of PAH molecules by old stellar populations, or a result of a poor correlation between $L_{\rm Br \alpha}$ and $L_{\rm [NeII]}$ for higher luminosities.
\item The nuclear spectrum of NGC~1386 shows mild silicate absorption at 9.7{\mm} and we estimate a silicate strength of $S_{\rm sil} = -0.69^{+0.19}_{-0.23}$. This feature is confined to the inner 20 pc, being 30\% weaker in the extended emission spectrum, implying that the silicate might be linked to the torus predicted by the AGN unified model. The same feature is found to be in emission in the Sy1 nucleus of NGC~7213 with $S_{\rm sil} = 0.15^{+0.14}_{-0.16}$. Both results agree with the unified model for AGN.
\item We employed the {\sc clumpy} torus models to further investigate the nuclear spectrum. Results for NGC~1386 show a column density of $N_{\rm H} = 1.4^{+0.1}_{-0.2} \times 10^{24}\,\,{\rm cm^{-2}}$ which is consistent with X-Ray estimates and favours the Compton thick scenario. For NGC~7213 the number of clouds in the line of sight is practically zero, with an estimated $N_{\rm H} \sim 10^{18}\,\,{\rm cm^{-2}}$, in agreement with its Sy1 classification.
\end{enumerate}

\section*{Acknowledgements}
We would link to thank an anonymous referee for the many suggestions which greatly improved the quality of the present paper. This work is based on observations obtained at the Gemini Observatory, which is operated by the Association of Universities for Research in Astronomy, Inc., under a cooperative agreement with the NSF on behalf of the Gemini partnership: the National Science Foundation (United States), the National Research Council (Canada), CONICYT (Chile), the Australian Research Council (Australia), Minist\'{e}rio da Ci\^{e}ncia, Tecnologia e Inova\c{c}\~{a}o (Brazil) and Ministerio de Ciencia, Tecnolog\'{i}a e Innovaci\'{o}n Productiva (Argentina). DRD thanks CNPq for financial support during this project. RR thanks CNPq and FAPERGS for financial support during this project.

\bibliographystyle{mn2e}
\bibliography{library}{}

\label{lastpage}

\end{document}